\documentclass[twocolumn]{aastex7}

\usepackage{CJK}
\usepackage{url}

\shorttitle{Quasar Radiative Feedback}
\shortauthors{Zhu et al.}

\begin{document}

\title{Quasar Radiative Feedback May Suppress Galaxy Growth on Intergalactic Scales at $z = 6.3$}

\author[0000-0003-3307-7525]{Yongda Zhu} 
\thanks{JASPER Scholar}
\affiliation{Steward Observatory, University of Arizona, 933 North Cherry Avenue, Tucson, AZ 85721, USA}
\email[show]{yongdaz@arizona.edu}


\author[0000-0003-1344-9475]{Eiichi Egami}
\affiliation{Steward Observatory, University of Arizona, 933 North Cherry Avenue, Tucson, AZ 85721, USA}
\email{}

\author[0000-0003-3310-0131]{Xiaohui Fan}
\affiliation{Steward Observatory, University of Arizona, 933 North Cherry Avenue, Tucson, AZ 85721, USA}
\email{}

\author[0000-0002-4622-6617]{Fengwu Sun}
\affiliation{Center for Astrophysics $|$ Harvard \& Smithsonian, 60 Garden St., Cambridge, MA 02138, USA}
\email{fengwu.sun@cfa.harvard.edu}


\author[0000-0003-2344-263X]{George D. Becker}
\affiliation{Department of Physics \& Astronomy, University of California, Riverside, CA 92521, USA}
\email{georgeb@ucr.edu}

\author[0000-0001-9420-7384]{Christopher Cain}
\affiliation{School of Earth and Space Exploration, Arizona State University, Tempe, AZ 85287-6004, USA}
\email{clcain3@asu.edu}

\author[0000-0002-3211-9642]{Huanqing Chen}
\affiliation{Augustana Campus,
University of Alberta,
Camrose, AB T4V2R3, Canada}
\email{huanqing.chen@ualberta.ca}

\author[0000-0003-2895-6218]{Anna-Christina Eilers}
\affiliation{MIT Kavli Institute for Astrophysics and Space Research, 77 Massachusetts Avenue, Cambridge, MA 02139, USA}
\affiliation{Department of Physics, Massachusetts Institute of Technology, Cambridge, MA 02139, USA}
\email{eilers@mit.edu}

\author[0000-0001-7440-8832]{Yoshinobu Fudamoto} 
\affiliation{Center for Frontier Science, Chiba University, 1-33 Yayoi-cho, Inage-ku, Chiba 263-8522, Japan}
\email{}

\author[0000-0003-4337-6211]{Jakob M. Helton}
\affiliation{Department of Astronomy \& Astrophysics, The Pennsylvania State University, University Park, PA 16802, USA}
\affiliation{Steward Observatory, University of Arizona, 933 North Cherry Avenue, Tucson, AZ 85721, USA}
\email{}

\author[0000-0002-5768-738X]{Xiangyu Jin}
\affiliation{Steward Observatory, University of Arizona, 933 North Cherry Avenue, Tucson, AZ 85721, USA}
\email{}

\author[0000-0003-4924-5941]{Maria Pudoka}
\affiliation{Steward Observatory, University of Arizona, 933 North Cherry Avenue, Tucson, AZ 85721, USA}
\email{}

\author[0000-0002-8651-9879]{Andrew J.\ Bunker}
\affiliation{Department of Physics, University of Oxford, Denys Wilkinson Building, Keble Road, Oxford OX1 3RH, UK}
\email{}

\author[0000-0001-8467-6478]{Zheng Cai}
\affiliation{Department of Astronomy, Tsinghua University, Beijing 100084, China}
\email{}

\author[0000-0002-6184-9097]{Jaclyn B. Champagne}
\thanks{JASPER Scholar}
\affiliation{Steward Observatory, University of Arizona,
933 North Cherry Avenue, Tucson, AZ 85721, USA}
\email{}

\author[0000-0001-7673-2257]{Zhiyuan Ji}
\affiliation{Steward Observatory, University of Arizona, 933 North Cherry Avenue, Tucson, AZ 85721, USA}
\email{}

\author[0000-0001-6052-4234]{Xiaojing Lin}
\affiliation{Department of Astronomy, Tsinghua University, Beijing 100084, China}
\email{}

\author[0000-0003-3762-7344]{Weizhe Liu}
\thanks{JASPER Scholar}
\affiliation{Steward Observatory, University of Arizona, 933 North Cherry Avenue, Tucson, AZ 85721, USA}
\email{}

\author[0000-0002-5237-9433]{Hai-Xia Ma}
\affiliation{Division of Particle and Astrophysical Science, Nagoya University, Furo-cho, Chikusa-ku, Nagoya 464-8602, Japan}
\email{}

\author[0009-0003-5402-4809]{Zheng Ma}
\affiliation{Steward Observatory, University of Arizona, 933 North Cherry Avenue, Tucson, AZ 85721, USA}
\email{}

\author[0000-0002-4985-3819]{Roberto Maiolino}
\affiliation{Kavli Institute for Cosmology, University of Cambridge, Madingley Road, Cambridge, CB3 OHA, UK}
\affiliation{Cavendish Laboratory - Astrophysics Group, University of Cambridge, 19 JJ Thomson Avenue, Cambridge, CB3 OHE, UK}
\affiliation{Department of Physics and Astronomy, University College London, Gower Street, London WC1E 6BT, UK}
\email{}

\author[0000-0003-2303-6519]{George H.\ Rieke}
\affiliation{Steward Observatory, University of Arizona, 933 North Cherry Avenue, Tucson, AZ 85721, USA}
\email{}

\author[0000-0002-7893-6170]{Marcia J.\ Rieke}
\affiliation{Steward Observatory, University of Arizona, 933 North Cherry Avenue, Tucson, AZ 85721, USA}
\email{}

\author[0000-0002-5104-8245]{Pierluigi Rinaldi}
\affiliation{AURA for the European Space Agency (ESA), Space Telescope Science Institute, 3700 San Martin Dr., Baltimore, MD 21218, USA}
\affiliation{Steward Observatory, University of Arizona, 933 North Cherry Avenue, Tucson, AZ 85721, USA}
\email{}

\author[0000-0001-6561-9443]{Yang Sun}
\affiliation{Steward Observatory, University of Arizona, 933 North Cherry Avenue, Tucson, AZ 85721, USA}
\email{}

\author[0000-0003-0747-1780]{Wei Leong Tee}
\affiliation{Steward Observatory, University of Arizona, 933 North Cherry Avenue, Tucson, AZ 85721, USA}
\email{}

\author[0000-0002-7633-431X]{Feige Wang}
\affiliation{Department of Astronomy, University of Michigan, 1085 S. University Ave., Ann Arbor, MI 48109, USA}
\email{}

\author[0000-0001-5287-4242]{Jinyi Yang}
\affiliation{Department of Astronomy, University of Michigan, 1085 S. University Ave., Ann Arbor, MI 48109, USA}
\email{}

\author[0000-0002-5367-8021]{Minghao Yue}
\affiliation{Steward Observatory, University of Arizona, 933 North Cherry Avenue, Tucson, AZ 85721, USA}
\affiliation{MIT Kavli Institute for Astrophysics and Space Research, 77 Massachusetts Avenue, Cambridge, MA 02139, USA}
\email{}

\author[0000-0002-1574-2045]{Junyu Zhang}
\affiliation{Steward Observatory, University of Arizona, 933 North Cherry Avenue, Tucson, AZ 85721, USA}
\email{}

\begin{abstract}
We present observational evidence that intense ionizing radiation from a luminous quasar suppresses nebular emission in nearby galaxies on intergalactic scales at $z=6.3$. Using JWST/NIRCam grism spectroscopy from the SAPPHIRES and EIGER programs, we identify a \added{moderate but statistically significant decline} in [O\,\textsc{iii}]\,$\lambda5008$ luminosity relative to the UV continuum ($L_{5008}/L_{1500}$) among galaxies within $\sim$\added{7} comoving Mpc (cMpc) of the quasar J0100$+$2802, the most UV-luminous quasar known at this epoch ($M_{1450}=-29.26$). While $L_{1500}$ remains roughly constant with transverse distance, $L_{5008}$ increases significantly, suggesting suppression of very recent star formation toward the quasar. The effect persists after controlling for completeness, local density, and UV luminosity, and correlates with the projected photoionization-rate profile $\Gamma_{\mathrm{qso}}$. A weaker but directionally consistent suppression in $L_{5008}/L_{1500}$ is also observed along the line of sight. The transverse suppression radius ($\sim$\added{7} cMpc) implies a recent radiative episode with a cumulative duration $\sim$\added{3.1 Myr}, shorter than required for thermal photoheating to dominate and thus more naturally explained by rapid H$_2$ photodissociation and related radiative processes. Environmental effects alone appear insufficient to explain the signal. Our results provide direct, geometry-based constraints on large-scale quasar radiative feedback and recent quasar lifetimes.
\end{abstract}


\section{Introduction} \label{sec:intro}

During the epoch of reionization, luminous quasars (QSOs) serve as natural laboratories for studying the interplay between intense radiation fields and early galaxy formation \citep[e.g.,][]{fan_quasars_2023, wang_luminous_2021, bosman_mature_2024}. These quasars illuminate their surroundings over intergalactic ($\sim$\,cMpc) scales \citep[e.g.,][]{eilers_detecting_2020}, raising a key question: can quasar radiation directly suppress star formation and metal-line cooling in nearby galaxies, beyond what is expected from standard environmental clustering?

Reionization-era quasars could be the signposts of protoclusters. Observationally, galaxies in dense environments, even at high redshift, exhibit accelerated evolution and more diverse properties \citep{dressler_galaxy_1980, helton_jwst_2024, helton_identification_2024, morishita_accelerated_2025, li_epochs_2025, witten_before_2025}. On average, protoclusters at $z \sim 2$--3 host more evolved, massive galaxies than the field, supporting an ``environmental bias'' scenario \citep[e.g.,][]{steidel_spectroscopic_2005, harikane_silverrush_2019, lin_luminosity_2025}. Galaxies at cosmic noon ($z\sim2-3$) located in high-density regions also tend to show elevated ionizing photon production efficiency \citep[][see also \citealp{li_reionization_2025}]{zhu_smiles_2025}. At higher redshift ($z \sim 6$), prior to the launch of JWST, quasar environments were primarily studied using Lyman-$\alpha$ emitters (LAEs) and Lyman-break galaxies (LBGs) \citep[e.g.,][]{kashikawa_habitat_2007, kim_environments_2009, utsumi_large_2010, banados_galaxy_2013, goto_no_2017, kikuta_active_2017, mazzucchelli_no_2017, ota_large-scale_2018, champagne_mixture_2023, pudoka_large-scale_2024, yue_escape_2025}. Both tracers are sensitive to environment, but LAEs are additionally affected by radiative transfer and residual neutral hydrogen \citep{dijkstra_saas-fee_2017}. Interestingly, several lower-redshift studies have found that LAEs either avoid overdense regions \citep{francis_mysterious_2004, kashikawa_habitat_2007, huang_evaluating_2022} or tend to populate lower-density environments \citep{cooke_nurturing_2013, witstok_inside_2024, witstok_jades_2025}. This behavior has been attributed to suppression of star formation in low-mass halos by an enhanced ultraviolet background \citep[UVB;][]{kashikawa_habitat_2007, bruns_clustering_2012, bosman_three_2020}, and to modifications in the circumgalactic medium (CGM) structure in dense environments where the CGM is truncated or stripped \citep{yoon_influence_2013}.

The launch of JWST \citep{gardner_james_2006}, and particularly the NIRCam \citep{rieke_performance_2023} grism mode \citep[e.g.,][]{greene__2017, sun_first_2023}, enables a robust alternative for measuring galaxy density via rest-frame optical emission lines, especially [O\,\textsc{iii}]\,$\lambda5008$, which is largely immune to Ly$\alpha$ radiative transfer effects \citep[e.g.,][]{osterbrock_astrophysics_2006, helton_jwst_2023, champagne_quasar-anchored_2025, champagne_quasar-anchored_2025-1, jin_spectroscopic_2024, fudamoto_sapphires_2025, stone_z708_2025}. Radiation-hydrodynamic simulations of massive halos at $z \sim 6$--7 predict significant impacts on nearby galaxy populations via quasar winds and radiation \citep{costa_hidden_2019, kim_high-redshift_2019, ni_gas_2018}. On larger scales, cosmological radiative transfer simulations quantify quasar radiative feedback and show that intense proximity-zone illumination can suppress star formation in small halos out to $\sim$10\,cMpc by photodissociating molecular hydrogen (H$_2$) and heating the gas \citep{chen_role_2020}.

In this Letter, we investigate whether high-redshift quasars can suppress star formation in neighboring galaxies through radiative feedback beyond their host halos, using a rest-frame optical tracer for the first time. We focus on J0100$+$2802 \citep{wu_ultraluminous_2015} at $z=6.3$ ($M_{1450}=-29.26$), the most UV-luminous quasar known at $z>6$. This quasar hosts a large Ly$\alpha$ proximity zone and lies within a large-scale overdensity traced by LBGs and LAEs \citep[e.g.,][]{eilers_implications_2017,pudoka_large-scale_2024,hashemi_ly_2025}. Using deep JWST/NIRCam imaging and grism spectroscopy from the Slitless Areal Pure-Parallel High-Redshift Emission Survey (SAPPHIRES; \citealp{sun_slitless_2025}) and Emission-line galaxies and Intergalactic Gas in the Epoch of Reionization (EIGER; \citealp{kashino_eiger_2022}) surveys, we identify [O\,\textsc{iii}]-emitting galaxies within approximately 20\,cMpc of the quasar.

The remainder of this Letter is organized as follows: Section~\ref{sec:data} describes the observations; Section~\ref{sec:results} presents the main results and tests; Section~\ref{sec:implications} discusses implications for quasar radiative feedback and lifetimes; Section~\ref{sec:conclusions} concludes. Throughout this letter, we adopt the {\tt Planck18} cosmology \citep{planck_collaboration_planck_2020} and quote comoving distances unless otherwise noted.

\section{Observations and Analysis} \label{sec:data}

\begin{figure*}[!ht]
\centering
\includegraphics[width=0.6\linewidth]{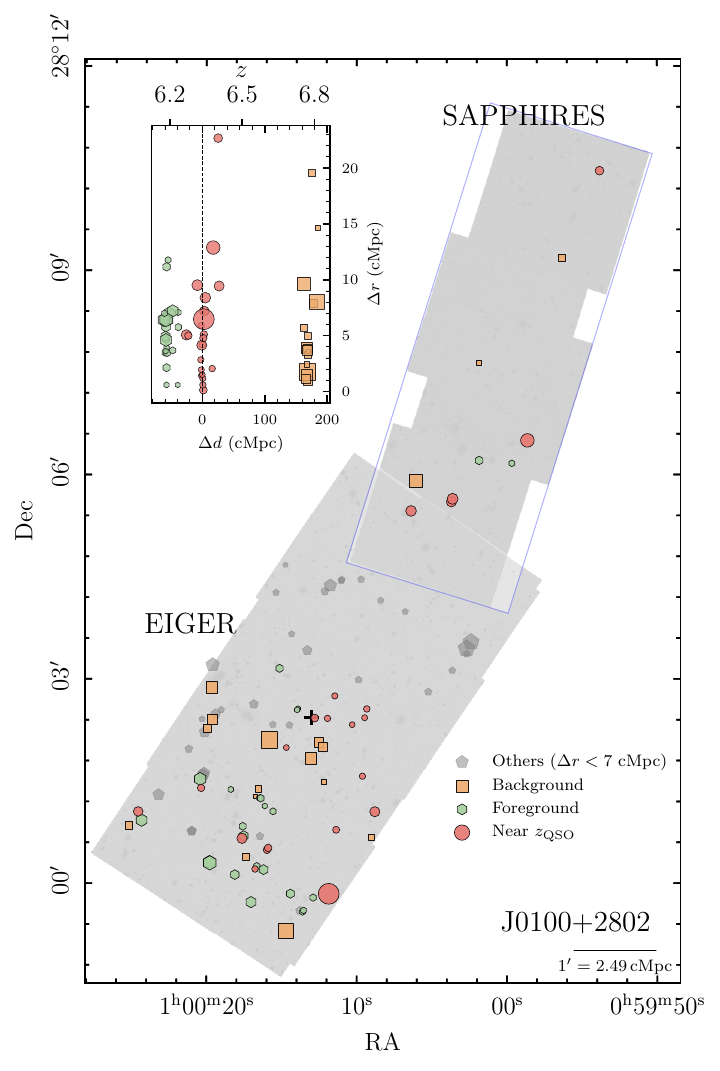}
\caption{
Sky map of the J0100$+$2802 field, combining NIRCam F356W mosaics from SAPPHIRES (top) and EIGER (bottom). Galaxies at $z \approx 6.3$ (near the quasar redshift; red circles), $z \approx 6.2$ (foreground; green hexagon), and $z \approx 6.8$ (background overdensity; orange squares) are shown with symbol sizes scaled by their observed [O\,\textsc{iii}]~$\lambda5008$ luminosities. The black cross marks the quasar position. 
We also show other galaxies with transverse distance from the quasar $\Delta r < 7$ cMpc on the sky map just for reference. The inset shows the projected distribution along the line of sight, with transverse separation $\Delta r$ plotted against line-of-sight distance $\Delta d$ from the quasar.
}
\label{fig:sky}
\end{figure*}

\subsection{JWST Imaging and Spectroscopy}

We use deep JWST/NIRCam imaging and grism spectroscopy of the $z = 6.3$ quasar J0100$+$2802 from the SAPPHIRES program (PID 6434; PI: E.~Egami) and the EIGER survey (PID 1243; PI: S.~J.~Lilly). 
\footnote{\added{All the JWST data used in this paper can be found in MAST: \dataset[https://doi.org/10.17909/qef7-vp09]{https://doi.org/10.17909/qef7-vp09}.}}
J0100$+$2802 is among the most UV-luminous quasars known at $z>6$ ($M_{1450} = -29.26$; \citealp{wu_ultraluminous_2015}), making it an ideal target for investigating radiative feedback on intergalactic scales. The SAPPHIRES observations were obtained in parallel to a NIRSpec \citep{jakobsen_near-infrared_2022} multi-object spectroscopy (MOS; \citealp{ferruit_near-infrared_2022}) observation from program 4713 (PI: A.-C.~Eilers) and consist of two overlapping NIRCam WFSS tiles. Each tile includes three dithered integrations with the GRISMR grism and F356W filter (paired with F200W for direct imaging), totaling 13,335\,s per tile. These were followed by three dithered exposures in direct imaging mode using F115W and F356W (CLEAR), also totaling 13,335\,s per tile. The combined on-source exposure time is 7.4\,hours per filter and mode, including F356W/GRISMR, F356W imaging, F200W, and F115W. EIGER observations centered on the quasar (see \citealt{kashino_eiger_2022}) were independently reprocessed and included in our analysis to increase the effective survey area and enable direct comparisons between quasar environments. In EIGER, the nominal exposure time in a single tile (visit) is 4380 seconds per filter for the short-wavelength (SW) imaging and 8760 seconds for the WFSS. 

We reduced all imaging data using a custom pipeline based on the JWST Calibration Pipeline (v1.11.3), with additional corrections for instrumental artifacts including $1/f$ noise and snowballs. WFSS reductions were performed using the open-source pipeline from \citet{sun_first_2023},\footnote{\url{https://github.com/fengwusun/nircam_grism}} with updated F356W trace and wavelength calibrations (see \citealp{sun_slitless_2025}). For each exposure we generated 2D dispersed spectra for all cataloged sources, accounting for geometric distortion and spectral tilt, interpolated them onto a common wavelength-spatial grid, and stacked. One-dimensional spectra were extracted with a fixed 5-pixel (0\farcs3) boxcar. Background and continuum were removed locally using median-filter windows. Emission-line candidates were identified by visual inspection and confirmed by the [O\,\textsc{iii}]$\lambda\lambda4960,5008$ doublet: the $\lambda5008$ line must be detected at $\geq 5\sigma$ with a corresponding $\lambda4960$ detection at $\geq 2\sigma$ at the expected separation. Redshifts were measured from [O\,\textsc{iii}]$\lambda5008$.

We define three redshift subsamples within the J0100 field: a proximity sample within $|\Delta z|\lesssim 0.07$ of the quasar redshift ($z\approx6.3$), a \emph{foreground} control at $z\approx6.2$ (about 60\,cMpc toward lower redshift), and a \emph{background} control at $z\approx6.8$ (about 170\,cMpc toward higher redshift). Because the observed [O\,\textsc{iii}] lines fall within a narrow wavelength range in F356W over $z\approx6.2$-6.8, the depth across the three slices is very similar. After quality cuts and artifact removal, the final catalog contains 24 galaxies in the foreground sample, 22 near $z\approx6.3$, and 17 in the background sample, out of 130 emitters over $5.3<z<6.9$. Figure~\ref{fig:sky} shows the SAPPHIRES and EIGER footprints with the detected [O\,\textsc{iii}] emitters in each slice.
\added{We note that the galaxies near the quasar are not azimuthally uniform on the sky, with most located to the south. The two-dimensional completeness map of the EIGER mosaics \citep{kashino_eiger_2025} shows nearly symmetric sensitivity across the field, and a similar excess of} [O\,\textsc{iii}] \added{emitters to the south is also seen in previous analyses of the J0100 field \citep{kashino_eiger_2022,kashino_eiger_2025}. This suggests that the observed anisotropy likely reflects genuine large-scale structure or cosmic variance rather than an observational bias.}

\subsection{Derived Quantities} \label{sec:analysis}

[O\,\textsc{iii}]$\lambda5008$ emission-line fluxes were measured from the one-dimensional, continuum-subtracted spectra by fitting a single Gaussian profile and integrating the model. Typical uncertainties on [O\,\textsc{iii}] fluxes are approximately 4\% based on propagated fitting errors. Rest-frame UV continuum luminosities ($L_{1500}$) were estimated from F115W photometry, assuming a flat $f_\nu$ spectrum. All fluxes were measured within a circular aperture of radius 0\farcs15, matching the spectral extraction aperture, using the \texttt{sep} package \citep{barbary_sep_2016}. We define the ratio $\log_{10}(L_{5008} / L_{1500})$ as a proxy for the ionized-gas cooling efficiency normalized by star formation rate (see Appendix~\ref{sec:robustness} for results based on $\log_{10}(L_{5008})$). Uncertainties on this ratio are dominated by the [O\,\textsc{iii}] flux error; the contribution from F115W photometric uncertainty is negligible.

To interpret the observed trends in [O\,\textsc{iii}]/UV ratios in the context of quasar radiative feedback, we model the local ionizing radiation field using a physically motivated prescription for the quasar photoionization rate $\Gamma_{\mathrm{qso}}(r)$. Rather than assuming pure geometric dilution, we follow the formalism developed by \citet{becker_mean_2021} and implemented in \citet{zhu_probing_2023}, in which attenuation of Lyman-continuum (LyC) photons by the IGM is included via a locally modified opacity $\kappa_{\rm LL}(r)$. This relation is based on stacked rest-UV spectroscopic observations of bright high-$z$ quasars from ground-based telescopes \citep[see also][]{prochaska_direct_2009, calverley_measurements_2011}. The opacity is parameterized as a function of total ionizing flux as:
\begin{equation}
\kappa_{\rm LL}(r) = \kappa_{\rm LL}^{\rm bg} \left[1 + \frac{\Gamma_{\mathrm{qso}}(r)}{\Gamma_{\mathrm{bg}}} \right]^{-\xi},
\end{equation}
where $\Gamma_{\mathrm{bg}}$ is the background photoionization rate and $\xi \approx 0.67$ is the power-law index describing how LyC opacity scales with the radiation field intensity \added{(see discussions on $\xi$ in \citealt{becker_mean_2021} and \citealt{zhu_probing_2023})}. This model captures the reduced opacity within the quasar proximity zone due to enhanced ionizing flux.

To compute $\Gamma_{\mathrm{qso}}(r)$, we adopt an ionizing spectral energy distribution (SED) with a power-law index of $\alpha_{\nu}^{\rm ion} = 1.5$ and normalize the quasar luminosity at 1450\,\AA\ to its observed $M_{1450}$. We then calculate the ionizing luminosity at 912\,\AA, $L_{912}$, assuming a broken power-law SED with $\alpha_{\nu}^{\rm UV} = 0.6$ for $\lambda > 912$\,\AA\ (see \citealp{becker_mean_2021} and references therein). The geometrically attenuated profile of $\Gamma_{\mathrm{qso}}(r)$ is computed iteratively in radial steps $\delta r$, starting with:
\begin{equation}
\Gamma_{\mathrm{qso}}(\delta r) = \Gamma_{\mathrm{bg}} \left( \frac{\delta r}{R_{\rm eq}} \right)^{-2},
\end{equation}
and propagating outward via:
\begin{equation}
\Gamma_{\mathrm{qso}}(r + \delta r) = \Gamma_{\mathrm{qso}}(r) \left( \frac{r + \delta r}{r} \right)^{-2} e^{-\kappa_{\rm LL}(r)\, \delta r},
\end{equation}
where $R_{\rm eq}$ is the characteristic distance at which $\Gamma_{\mathrm{qso}} = \Gamma_{\mathrm{bg}}$ in the absence of attenuation, given by:
\begin{equation}
R_{\rm eq} = \left[ \frac{L_{912} \sigma_0}{8\pi^2 \hbar \Gamma_{\mathrm{bg}}(\alpha_{\nu}^{\rm ion} + 2.75)} \right]^{1/2},
\end{equation}
as defined in \citet{calverley_measurements_2011}. Here, $\sigma_0$ is the hydrogen ionization cross section at 912\,\AA, $\hbar$ is the reduced Planck constant, and $\alpha_{\nu}^{\rm ion}$ is the spectral slope of the quasar ionizing continuum (for $\lambda < 912$\,\AA) in the frequency domain. We adopt $\Gamma_{\mathrm{bg}} = 2 \times 10^{-13}\,\mathrm{s^{-1}}$ at $z = 6.3$, based on recent constraints from \citet{gaikwad_measuring_2023} \citep[see also][]{davies_constraints_2023}. This model enables a more accurate representation of quasar radiation profiles on intergalactic scales, which we compare to observed trends in [O\,\textsc{iii}]/UV ratios as a function of distance.

Finally, we assess the detection completeness of [O\,\textsc{iii}] sources as a joint function of $L_{5008}$ and transverse distance from the quasar. We inject synthetic line emitters into the 3D error cubes over a grid in redshift and sky position, assuming a fixed line width of 350\,km\,s$^{-1}$ and a range of line fluxes, and recover them with the same pipeline (see \citealp{zhu_galaxy_2025}). The resulting completeness map for the J0100 field is consistent with \citet{kashino_eiger_2025}. Completeness is computed as a function of $L_{5008}$ and sky position and then averaged in bins of transverse distance from the quasar $\Delta r$. We adopt $L_{5008}>10^{42}\,\mathrm{erg\,s^{-1}}$ as a detection threshold, for which completeness is at least 40\% in both the EIGER and SAPPHIRES fields. For the science analyses we further require $>$50\% completeness within each radial bin, and for bootstrap tests we downsample to a uniform 60\% completeness. Only galaxies in regions meeting these criteria are included in the trend analyses and regression fits; in the $z\approx6.3$ sample, 20 of 22 sources pass these cuts. Raising the local completeness requirement to 75\% yields the same qualitative suppression in $L_{5008}$.

\begin{figure*}[!ht]
\centering
\includegraphics[width=0.48\linewidth]{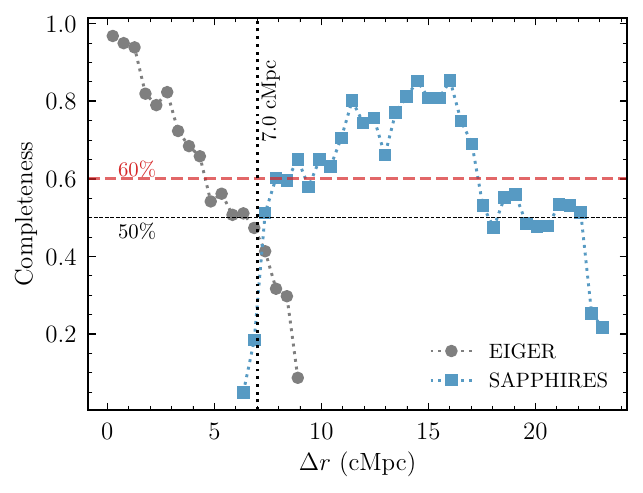}
\includegraphics[width=0.49\linewidth]{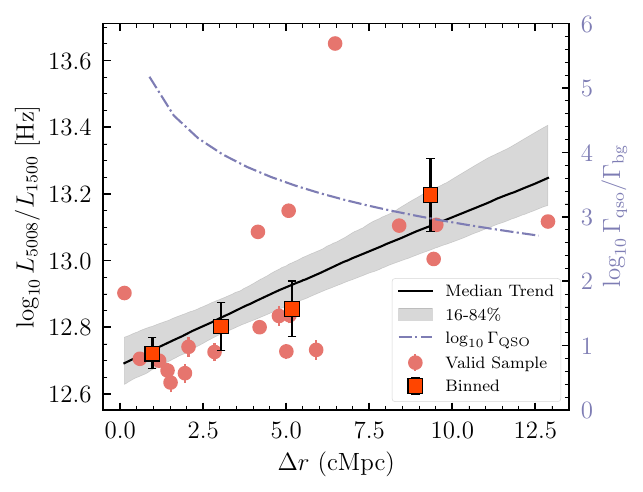}
\caption{
\textbf{Left:} Field-averaged detection completeness for galaxies with $L_{5008} > 10^{42}\,\mathrm{erg\,s^{-1}}$ as a function of projected transverse distance $\Delta r$ (cMpc) from J0100$+$2802, for the SAPPHIRES (blue squares) and EIGER (gray circles) mosaics. The horizontal dashed lines mark 50\% and 60\% completeness. The vertical dotted line at $\Delta r = 7$ cMpc indicates where both fields reach $\gtrsim 50\%$ completeness (from injection-recovery tests). 
\textbf{Right:} [O\,\textsc{iii}]-to-UV luminosity ratio ($\log_{10}(L_{5008}/L_{1500})$) versus $\Delta r$ for galaxies that pass the $\Delta r$-dependent completeness cut ($>50\%$; red points). The solid black line is the best-fit linear trend \added{derived from the} \textit{unbinned data points}, and the shaded band shows the 16th to 84th percentile bootstrap interval from mock downsampling to uniform 60\% completeness. Binned averages (five galaxies per bin) are shown as red squares with error bars, \added{which represent 1$\sigma$ uncertainties in the mean within each bin}. The dot-dashed purple curve shows the expected quasar photoionization rate, and the right-hand axis gives $\log_{10}(\Gamma_{\rm QSO}/\Gamma_{\rm bg})$.
}
\label{fig:trend}
\end{figure*}

\begin{figure*}[!ht]
\centering
\includegraphics[width=0.49\linewidth]{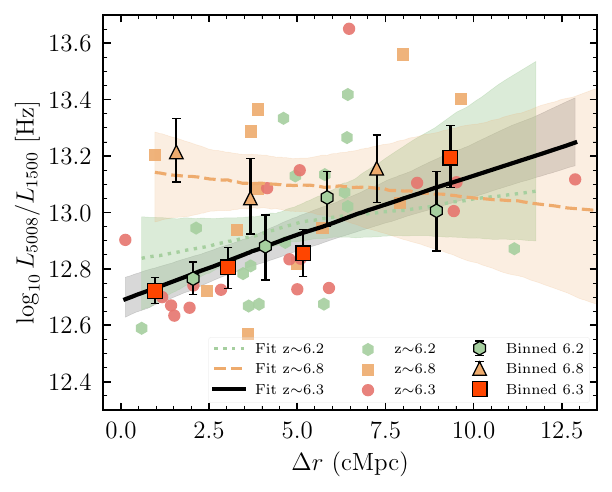}
\includegraphics[width=0.49\linewidth]{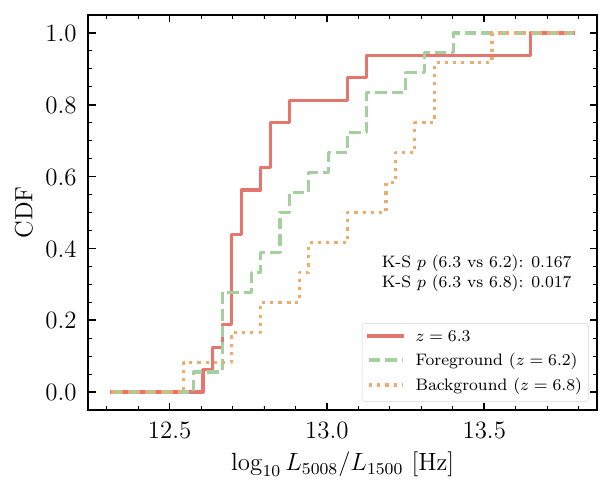}
\caption{
\textbf{Left:} Similar to Figure \ref{fig:trend}, [O\,\textsc{iii}]-to-UV ratio $\log_{10}(L_{5008}/L_{1500})$ versus transverse distance $\Delta r$ for galaxies near the quasar redshift ($z \approx 6.3$, red), in the foreground ($z \approx 6.2$, green), and in the background overdensity ($z \approx 6.8$, orange). Only data points with $>50\%$ completeness are shown here. Solid, dotted, and dashed curves show completeness-matched median fits for the three slices, with shaded bands indicating the 16--84\% ranges. Binned averages (five galaxies per bin) with bootstrap uncertainties are overplotted.
\textbf{Right:} Cumulative distributions (CDFs) of $\log_{10}(L_{5008}/L_{1500})$ for galaxies within $\Delta r < 7$\,cMpc of the quasar, comparing $z \approx 6.3$ (red), $z \approx 6.2$ (green), and $z \approx 6.8$ (orange). The background sample shows higher ratios than the $z \approx 6.3$ sample (K--S $p=0.017$), while the foreground is more similar ($p=0.167$), consistent with the visual impression from the left panel.
}
\label{fig:cdf}
\end{figure*}

\section{Results and Discussion}  \label{sec:results}

\subsection{Suppression with Transverse Distance}
\label{sec:trend}

Figure~\ref{fig:trend} (left) shows the field-averaged detection completeness for galaxies with $L_{5008} > 10^{42}\,\mathrm{erg\,s^{-1}}$ as a function of projected transverse distance $\Delta r$ from J0100$+$2802. Completeness peaks near the center of the EIGER mosaic, declines toward its edges where exposure time is lower, and rises again in the overlap region near the center of the SAPPHIRES field. The vertical dotted line at $\Delta r = 7$\,cMpc marks the approximate transition between the two mosaics, where both reach about 50\% completeness. To assess systematics, we also perform forward modeling of the observed [O\,\textsc{iii}] luminosity distribution by convolving a published [O\,\textsc{iii}] luminosity function with the position-dependent completeness; results are shown in Appendix~\ref{sec:robustness} and confirm that the measured trends are not driven by completeness gradients. Here we present the empirical completeness profiles used for the main selection and fits.

In the right panel of Figure~\ref{fig:trend}, we present the central result of this study: \added{a moderate but statistically significant} decrease in $\log_{10}(L_{5008}/L_{1500})$ with decreasing projected distance from the quasar, consistent with radiative feedback that suppresses recent star formation (see Section~\ref{sec:implications}). Red points indicate individual galaxies that meet a local completeness threshold of $>50\%$ at their respective $\Delta r$; only these galaxies are included in the regression analysis. Binned averages (five galaxies per bin) are shown as orange squares with error bars and are broadly consistent with the fitted trend. \added{The error bars on the binned points represent 1$\sigma$ uncertainties in the mean within each bin.} The solid black line indicates the best-fit linear trend, with the shaded region representing the 16th to 84th percentile range from 500 bootstrap realizations. \added{The fitting is performed on the unbinned data points, while the binned averages are shown only for visual reference.} Each bootstrap draws a completeness-matched mock sample assuming a uniform 60\% detection probability. We find a slope of $\log_{10}(L_{5008}/L_{1500})$-$\Delta r$ of \added{$0.043^{+0.016}_{-0.008}$ and a Pearson $p$-value of 0.012, indicating $>98\%$ significance}. Given the limited radial range probed and sky coverage, we refer to this as a \emph{trend} over the observed scales rather than a universal relation. \added{When the single high-ratio source at $\Delta r \sim 7$\,cMpc is excluded, the fitted slope remains positive within 1$\sigma$ uncertainty ($0.039^{+0.007}_{-0.009}$, $p=0.015$), confirming that the overall trend is not driven by one outlier. For consistency with the completeness transition and the region of strongest suppression, we therefore adopt a characteristic radius of $\sim7$\,cMpc to describe the observed effect.}

The dot-dashed purple curve in Figure~\ref{fig:trend} shows the predicted photoionization rate $\Gamma_{\mathrm{qso}}$ as a function of distance from the quasar, normalized by the metagalactic ionizing background $\Gamma_{\mathrm{bg}}$. The observed trend in $\log_{10}(L_{5008}/L_{1500})$ increases with distance from the quasar, in contrast to the declining $\Gamma_{\mathrm{qso}}(r)$ profile. A more detailed discussion of the physical interpretation is deferred to Section~\ref{sec:implications}.

\subsection{Line-of-Sight Trends in [O\,\textsc{iii}] Luminosity} \label{sec:density}

To assess whether quasar radiative feedback manifests along the line of sight (LOS), we compare galaxies at similar projected separations but offset in redshift. Figure~\ref{fig:cdf} now shows both the transverse trends and the LOS comparison: the left panel plots $\log_{10}(L_{5008}/L_{1500})$ versus $\Delta r$ for the near-$z_{\rm QSO}$ slice ($z\approx6.3$; red), the foreground ($z\approx6.2$; green), and the background overdensity ($z\approx6.8$; orange), with completeness-matched fits and binned averages; the right panel presents cumulative distributions within $\Delta r<7$\,cMpc.

The background sample lies systematically above the $z\approx6.3$ sequence at fixed $\Delta r$ and shows only a weak slope, whereas the $z\approx6.3$ slice exhibits the strong positive trend with distance seen in Section~\ref{sec:trend}. Within $\Delta r<7$\,cMpc, the CDFs confirm this offset: the background ratios are higher than those at $z\approx6.3$ (K--S $p=0.017$). By contrast, the foreground slice overlaps the $z\approx6.3$ sample in both normalization and slope, and its CDF is statistically consistent with $z\approx6.3$ (K--S $p=0.167$). A similar behavior is seen for $L_{5008}$ alone (K--S $p=0.003$ for $z\approx6.8$ vs.\ $z\approx6.3$, and $p=0.332$ for $z\approx6.2$ vs.\ $z\approx6.3$).

A natural explanation is that the $z=6.2$ slice resides within the extended LOS proximity zone of J0100$+$2802, which spans tens of cMpc due to light-travel delays and the high intrinsic luminosity of the quasar \citep[e.g.,][]{eilers_implications_2017}. In this geometry the LOS is effectively observed at a single quasar-on time, so a uniform suppression within the proximity zone would produce little residual correlation with LOS distance, consistent with the data. The $z\approx6.2$ galaxies may therefore experience ionization conditions similar to those at $z\approx6.3$.

In contrast, transverse measurements are sensitive to finite light-travel time. The significant suppression with decreasing $\Delta r$ (Figure~\ref{fig:trend}) implies that not all galaxies at large projected separations have yet been reached by the ionization front. This spatial anisotropy suggests a relatively recent turn-on, with an age shorter than the light-crossing time across the transverse separation, making the transverse direction a more temporally localized probe of quasar radiative feedback.

\subsection{Independence from Environment and $L_{1500}$}\label{sec:control}

\begin{figure*}[!ht]
    \centering
    \includegraphics[width=0.49\linewidth]{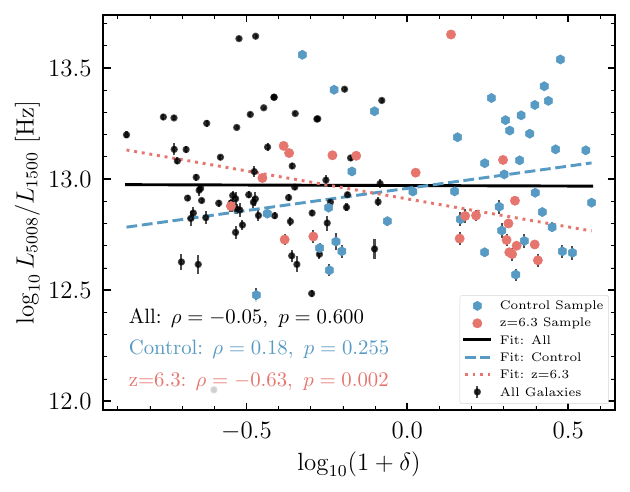}
    \includegraphics[width=0.49\linewidth]{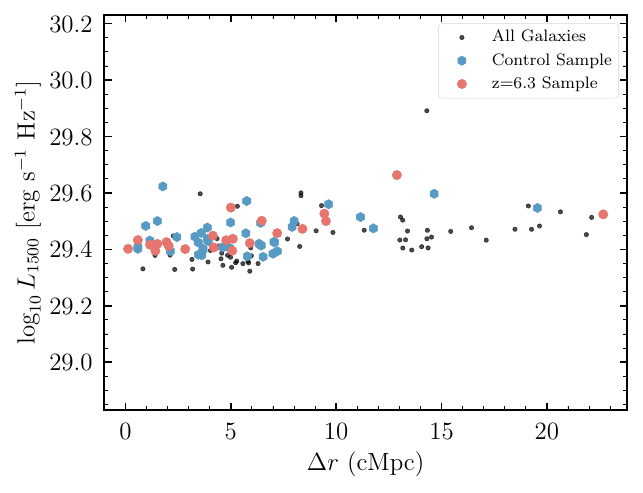}
\caption{
\textbf{Left:} [O\,\textsc{iii}]-to-UV luminosity ratio ($\log_{10}(L_{5008}/L_{1500})$) versus local overdensity, estimated with a $k=5$ nearest-neighbor metric in 3D over the full sample of 130 emitters ($5.3<z<6.9$) and normalized to the field average. The combined control sample (foreground $z\approx6.2$ plus background $z\approx6.8$; blue hexagons) shows a weak positive trend ($\rho=0.18$, $p=0.255$), while the $z\approx6.3$ near-QSO sample (red circles) shows a negative trend ($\rho=-0.63$, $p=0.002$). The full sample shows no significant correlation ($\rho=-0.05$, $p=0.600$). These patterns indicate that the quasar-distance dependence is unlikely to be driven by local overdensity alone.
\textbf{Right:} Rest-frame UV luminosity ($\log_{10} L_{1500}$) as a function of transverse distance $\Delta r$ from J0100$+$2802. Both the near-$z_{\rm QSO}$ and control samples show no significant trend with distance, with a typical scatter of about 0.2 dex. The y-axis span (in dex) matches Figure~\ref{fig:trend} (right) for visual comparison.
}
    \label{fig:controls}
\end{figure*}

To test whether the observed suppression in the [O\,\textsc{iii}]-to-UV luminosity ratio is driven by environmental effects or intrinsic galaxy properties, we compare $\log_{10}(L_{5008}/L_{1500})$ against two control variables: local overdensity and UV continuum luminosity.

Figure~\ref{fig:controls} (left) shows the [O\,\textsc{iii}]/UV ratio as a function of local overdensity, defined as $\log_{10}(1+\delta)$ using a $k=5$ nearest-neighbor estimator in 3D comoving space, applied to the full sample of 130 [O\,\textsc{iii}] emitters ($5.3<z<6.9$) across both fields. The overdensity $\delta$ is normalized by the field-average number density. We caution that this is a rough estimate because peculiar velocities are unknown and the measurements cover a modest field. We find no significant correlation between $\log_{10}(L_{5008}/L_{1500})$ and $\log_{10}(1+\delta)$ for the full sample (Spearman $\rho=-0.05$, $p=0.600$). The combined control sample (foreground $z\approx6.2$ plus background $z\approx6.8$; blue) shows a weak positive trend ($\rho=0.18$, $p=0.255$), whereas the near-$z_{\rm QSO}$ sample at $z\approx6.3$ (red) shows a negative trend ($\rho=-0.63$, $p=0.002$). Repeating this analysis with different $k$ values or using kernel density estimation (KDE) produces consistent results. The absence of a strong global trend with overdensity suggests that environment alone is unlikely to explain the drop in [O\,\textsc{iii}]/UV near the quasar, especially since galaxies in dense regions show diverse properties, as discussed in Section~\ref{sec:intro}.

In Figure~\ref{fig:controls} (right), we examine $L_{1500}$ as a function of transverse distance $\Delta r$ from J0100$+$2802. Both the near-$z_{\rm QSO}$ sample ($z\approx6.3$) and the combined control sample exhibit similar $\log_{10}(L_{1500})$ distributions (K-S $p$-value $>0.05$) with no significant radial gradient and a typical scatter of $\sim$0.2 dex. This confirms that the underlying UV luminosity is effectively matched across the samples.

Together, these control tests indicate that the observed decline in the [O\,\textsc{iii}]/UV ratio near the quasar is not driven by local overdensity or $L_{1500}$ variations (see, e.g., \citealp{pudoka_large-scale_2024, balmaverde_primordial_2017}). Instead, it primarily reflects a decline in [O\,\textsc{iii}] emission and points to a distinct mechanism that selectively suppresses nebular line emission while leaving the stellar continuum relatively unaffected in the vicinity of this ultra-luminous quasar.

\subsection{Cross-field Comparison of [O\textsc{iii}] Suppression and Quasar Ionizing Output}

Quasar radiative feedback is expected to depend on both the ionizing luminosity of the source and the distance to surrounding galaxies. To evaluate whether the [O\,\textsc{iii}] suppression observed near J0100$+$2802 is unique or part of a broader trend, we compare its impact with other luminous $z \approx 6.3$ quasars from the EIGER survey \citep{kashino_eiger_2025}, using our own uniform data reduction for consistency. The comparison sample comprises \object{J1148+5251} ($z_{\mathrm{qso}} = 6.422$, $M_{1450} = -27.62$; \citealp{banados_pan-starrs1_2016, shen_gemini_2019}), \object{J1030+0524} ($z_{\mathrm{qso}} = 6.304$, $M_{1450} = -26.99$; \citealp{dodorico_xqr-30_2023, mazzucchelli_xqr-30_2023}), and \object{J159$-$02} ($z_{\mathrm{qso}} = 6.381$, $M_{1450} = -26.47$; \citealp{banados_pan-starrs1_2016, farina_x-shooteralma_2022}).

\begin{figure*}[!ht]
\centering
\includegraphics[width=0.49\linewidth]{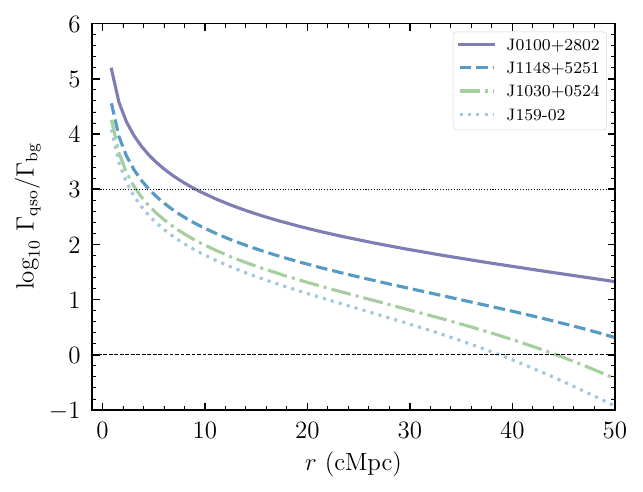}
\includegraphics[width=0.49\linewidth]{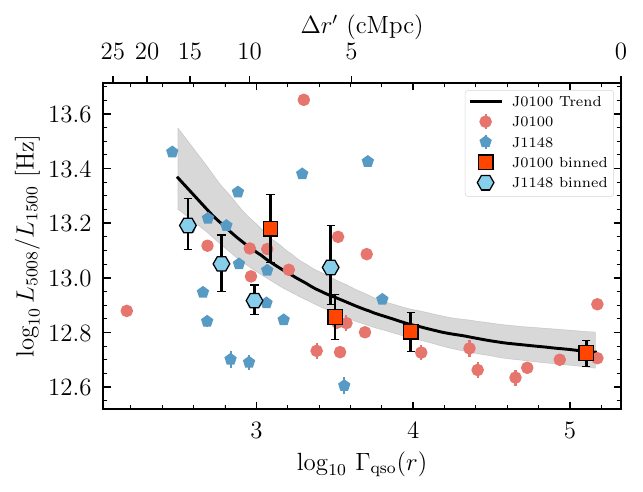}
\caption{
\textbf{Left:} Predicted quasar photoionization rate $\Gamma_{\rm QSO}$ relative to the metagalactic UV background $\Gamma_{\rm bg}$ as a function of comoving distance $r$ for four luminous quasars in the EIGER fields at $z \approx 6.3$. J0100$+$2802 (solid purple) exhibits the strongest radiative impact. The horizontal dashed and dotted lines indicate $\Gamma_{\rm QSO}/\Gamma_{\rm bg}=1$ and $10^3$. 
\textbf{Right:} [O\,\textsc{iii}]-to-UV luminosity ratio $\log_{10}(L_{5008}/L_{1500})$ versus $\log_{10}\Gamma_{\rm QSO}(r)$ for galaxies near J0100$+$2802 (red circles) and J1148$+$5251 (blue pentagons). Distances for J1148 sources are rescaled to match the J0100 photoionization profile; the top axis shows the corresponding scaled transverse separation $\Delta r'$ (cMpc). The black curve and shaded band show the best-fit relation and 68\% interval from the J0100 field in Figure~\ref{fig:trend} (right), re-plotted as a function of $\Gamma_{\rm QSO}(r)$. \added{The J1148 data show a qualitatively similar but weaker trend, with larger scatter likely due to smaller sample size and lower quasar luminosity.} Despite this, the J1148 binned trend (cyan hexagons) is broadly consistent with the J0100 binned points (orange squares). The other EIGER quasars (J1030$+$0524, J159$-$02) lack sufficient [O\,\textsc{iii}] emitters at small radii to assess this effect.
}
\label{fig:compare}
\end{figure*}

The left panel of Figure~\ref{fig:compare} shows the predicted quasar photoionization rate $\Gamma_{\mathrm{qso}}(r)$ normalized by the metagalactic UV background $\Gamma_{\mathrm{bg}}$ for four luminous quasars in EIGER. These profiles are computed using the attenuation-based formalism described in Section~\ref{sec:analysis}, which incorporates LyC photon absorption via a locally modified opacity \citep{becker_mean_2021, zhu_probing_2023}. Among the four, J0100$+$2802 (solid purple) exhibits the highest $\Gamma_{\mathrm{qso}}$, exceeding $\Gamma_{\mathrm{bg}}$ by over three orders of magnitude within $r \lesssim 7$\,cMpc. This corresponds closely to the observed suppression radius in $\log_{10}(L_{5008}/L_{1500})$, where values fall below the field median. J1148$+$5251 (dashed blue), while less luminous ($M_{1450} \approx -27.62$), still reaches $\Gamma_{\mathrm{qso}}/\Gamma_{\mathrm{bg}} \gtrsim 10^3$ within $\sim$5\,cMpc, allowing a meaningful comparison of feedback strength across quasar luminosities.

In the right panel of Figure~\ref{fig:compare}, we rescale the projected transverse distance $\Delta r$ for sources in the J1148$+$5251 field to a new variable $\Delta r'$, defined such that $\Gamma_{\mathrm{qso,\,J1148}}(\Delta r') = \Gamma_{\mathrm{qso,\,J0100}}(\Delta r)$. This enables a direct comparison of [O\,\textsc{iii}]-to-UV ratios at fixed ionizing exposure across fields of differing quasar luminosity.

We find that galaxies around J1148$+$5251 are broadly consistent with the suppression trend observed in the J0100 field, though with increased scatter. Notably, they do not show elevated $\log_{10}(L_{5008}/L_{1500})$ at fixed $\Gamma_{\mathrm{qso}}$, implying that less luminous quasars may still induce detectable feedback when $\Gamma_{\mathrm{qso}}/\Gamma_{\mathrm{bg}} \gtrsim 10^3$. In contrast, the fields around J1030$+$0524 and J159$-$02 lack bright [O\,\textsc{iii}] emitters at small scaled distances ($\Delta r' \lesssim 7$\,cMpc; see also \citealp{kashino_eiger_2025}), where suppression would be expected if strong feedback were present. Taken together, the scarcity of bright emitters at small scaled radii across these fields supports the interpretation that radiative suppression is genuine \citep[see also][]{champagne_quasar-anchored_2025, pudoka_lyman-break_2025}.

\section{Interpretation and Implications}
\label{sec:implications}

The observed decline in $\log_{10}(L_{5008}/L_{1500})$ with decreasing distance to J0100$+$2802, despite no corresponding trend in UV luminosity or in local galaxy density, suggests a mechanism that suppresses nebular line emission without immediately affecting the stellar continuum. The most plausible explanation is radiative feedback from the quasar acting on intergalactic scales, disrupting the ionization and cooling conditions needed for [O\,\textsc{iii}] excitation while leaving the legacy of recent star formation traced by $L_{1500}$ intact.

Radiation-hydrodynamic simulations by \citet{chen_role_2020} show that intense quasar radiation can rapidly suppress star formation in small galaxies, especially in low-mass halos that cannot self-shield, primarily through the photodissociation of H$_2$. This suppression occurs within approximately 1 physical cMpc of the AGN, with star formation rates reduced by around 0.5 dex in low-mass to average satellites, while more massive satellites remain less affected. Although line emission is not explicitly modeled, the associated delays in star formation are consistent with our finding that $L_{1500}$ remains stable while [O\,\textsc{iii}] declines near J0100. This supports a scenario in which nebular cooling and recent star formation are suppressed without erasing the established stellar population.

Meanwhile, thermal photoheating operates over much longer timescales (greater than $10^7$ yr), gradually increasing ISM thermal pressure and suppressing cold gas accretion \citep{costa_hidden_2019}. Proximity-zone measurements for J0100 indicate a short \emph{recent} episode, $t_{\mathrm{Q}}\sim10^5$ yr \citep{eilers_implications_2017}. This is broadly consistent with a flickering scenario in which the cumulative on-time can reach a few Myr (as implied by the transverse suppression; Section~\ref{sec:trend}), yet it remains far shorter than the $\gtrsim10^7$ yr required for thermal photoheating to dominate. Instead, more immediate radiative effects, such as H$_2$ photodissociation by Lyman-Werner photons, are more plausible. These can suppress dense-gas formation and nebular emission on short timescales while preserving the UV-bright stellar component, consistent with analytic models of abrupt ISM phase transitions triggered by critical ionizing backgrounds \citep{sternberg_h_2014}.

The transition in $\log_{10}(L_{5008}/L_{1500})$ spans nearly one dex in $\Gamma_{\mathrm{qso}}$, potentially indicating a nonlinear, threshold-like response rather than a smooth scaling with ionizing flux. Alternatively, the apparent jump between the 5 and 8 cMpc bins could reflect a geometric effect, such as a drop in $\Gamma_{\mathrm{qso}}$ due to shielding by a residual neutral patch. However, such structures are expected to disappear quickly given the brightness of the quasar, unless they lie outside its ionization cone. Combined with the absence of a correlation between [O\,\textsc{iii}] suppression and galaxy overdensity (Figure~\ref{fig:controls}), these results favor a direct radiative origin. That said, environmental mechanisms such as tidal interactions, ram pressure, or halo-related quenching could also modulate the specific star formation rate and line emission near the quasar (see Section~\ref{sec:intro}). While difficult to isolate with current statistics, future surveys with larger fields and deeper spectroscopic coverage will be essential to disentangle radiative and environmental effects.


The J1148$+$5251 field \added{shows a qualitatively similar but weaker trend, with} $\log_{10}(L_{5008}/L_{1500})$ \added{values that scatter both above and below the J0100 relation. This larger scatter likely reflects a combination of smaller sample size and lower quasar luminosity. Nonetheless, the data remain consistent with mild} radiative suppression when $\Gamma_{\mathrm{qso}} / \Gamma_{\mathrm{bg}} \gtrsim 10^3$, \added{suggesting that such effects may extend to moderately luminous quasars near the end of reionization rather than being uniformly strong.} \added{Additional evidence comes from the absence of bright} [O\,\textsc{iii}] emitters near other luminous quasars in EIGER, particularly J1030$+$0524 and J159$-$02. These fields show a deficit of [O\,\textsc{iii}] sources within scaled distances $\Delta r' \lesssim 7$\,cMpc, where suppression would be expected if \added{strong} feedback were present. This could reflect \added{either genuine feedback-induced suppression or} a detection bias due to galaxies becoming too faint in [O\,\textsc{iii}] (i.e., $\log_{10}(L_{5008}) < 41.6$) to be identified in current observations \citep[see, e.g.,][]{stone_z708_2025}, possibly due to older stellar populations or low specific star formation rates. \added{In either case, the overall pattern across multiple sightlines is consistent with a widespread but variable form of radiative influence.}

Finally, the \added{moderate but statistically significant} suppression of [O\,\textsc{iii}] at $\Delta r \lesssim 7$ cMpc \added{in the J0100 field} corresponds to a light-travel distance of approximately \added{1} proper Mpc, implying a lower bound on the quasar's radiative lifetime of about \added{3.1 Myr}. This is significantly longer than typical estimates based on Ly$\alpha$ proximity-zone sizes and may indicate that the [O\,\textsc{iii}] suppression reflects the integrated impact of multiple episodic bursts, with the proximity zone tracing only the most recent active phase. The presence of an extended transverse suppression region, in contrast to a compact proximity zone, supports a scenario in which the quasar flickers on timescales shorter than the timescale on which [O\,\textsc{iii}] emission responds to photoionization. Because transverse distances are less affected by light-cone delays and IGM attenuation, this geometry enables a complementary and potentially more robust constraint on the cumulative radiative history of quasars. Future work may test this scenario by searching for systematic differences in ionization-sensitive line ratios between galaxies exposed to recent versus older quasar episodes. One related caveat is that intrinsic galaxy property gradients near the quasar, for example higher metallicity, could also weaken [O\,\textsc{iii}] at fixed sSFR; targeted NIRSpec MSA spectroscopy of diagnostic line ratios can test this.

\section{Conclusions} \label{sec:conclusions}

We present observational evidence that radiative feedback from the luminous $z=6.3$ quasar J0100$+$2802 suppresses [O\,\textsc{iii}] emission in nearby galaxies on intergalactic scales. Using deep JWST/NIRCam grism spectroscopy from the SAPPHIRES and EIGER programs, we find:

\begin{itemize}
    \item A \added{moderate but statistically significant} decline in $\log_{10}(L_{5008}/L_{1500})$ with decreasing transverse distance from the quasar, robust to variations in completeness, environment, and UV luminosity. The suppression radius of $\sim$\added{7}\,cMpc implies a recent radiative burst with a minimum duration of $\sim$\added{3.1}\,Myr from a light-travel time argument.
    
    \item A weaker trend along the line of sight, consistent with extended quasar influence but less pronounced, possibly due to anisotropic emission, IGM attenuation, or light-cone effects. If confirmed with a larger sample, the transverse suppression may provide a powerful constraint on quasar lifetimes and duty cycles.
    
    \item A comparison with other $z\sim6.3$ quasars shows that the suppression trend near J1148$+$5251 is broadly consistent with that of J0100, while the deficit of bright [O\,\textsc{iii}] emitters near J1030$+$0524 and J159$-$02 suggests that such suppression may be widespread. These findings imply that emission-line-selected surveys (e.g., [O\,\textsc{iii}]) may underestimate galaxy densities around luminous quasars.
\end{itemize}

These results are consistent with a scenario in which quasar radiation rapidly alters ISM conditions, likely through H$_2$ photodissociation or ionization, suppressing recent star formation without immediately affecting the UV-bright stellar population. JWST provides direct constraints on such radiative feedback during reionization, offering a window into quasar-galaxy interaction at the highest redshifts. Future wide-field NIRCam imaging and WFSS around luminous quasars will be essential for mapping the full extent of this suppression. NIRSpec follow-up can test this scenario by searching for weak nebular lines, low ionization parameters, or older stellar populations in continuum-selected galaxies near quasars.

\begin{acknowledgments}
\added{We thank the anonymous reviewer for their constructive feedback.}
Y.Z.\ and J.M.H.\ acknowledge support from the NIRCam Science Team contract to the University of Arizona, \added{NAS5-02105}. 
J.M.H.\ is also supported by JWST Program 3215.
G.D.B.\ is supported by JWST Program 4092.
Support for programs \#3215, \#4092, and \#6434 was provided by NASA through a grant from the Space Telescope Science Institute, which is operated by the Association of Universities for Research in Astronomy, Inc., under NASA contract NAS 5-03127.
H.C.\ thanks the support by the Natural Sciences and Engineering Research Council of Canada (NSERC), funding reference \#RGPIN-2025-04798 and \#DGECR-2025-00136, and by the University of Alberta, Augustana Campus.
C.C.\ was supported by the Beus Center for Cosmic Foundations. 
A.J.B.\ acknowledges funding from the ``FirstGalaxies'' Advanced Grant from the European Research Council (ERC) under the European Union’s Horizon 2020 research and innovation program (Grant agreement No. 789056).

This work is based on observations made with the NASA/ESA/CSA James Webb Space Telescope. The data were obtained from the Mikulski Archive for Space Telescopes at the Space Telescope Science Institute, which is operated by the Association of Universities for Research in Astronomy, Inc., under NASA contract NAS 5-03127 for JWST. These observations are associated with programs \#6434 and \#1243. \added{All of the data presented in this paper were obtained from the Mikulski Archive for Space Telescopes (MAST) at the Space Telescope Science Institute. The specific observations analyzed can be accessed via \dataset[https://doi.org/10.17909/qef7-vp09]{https://doi.org/10.17909/qef7-vp09}. STScI is operated by the Association of Universities for Research in Astronomy, Inc., under NASA contract NAS5-26555. Support to MAST for these data is provided by the NASA Office of Space Science via grant NAG5–7584 and by other grants and contracts.}

We respectfully acknowledge the University of Arizona is on the land and territories of Indigenous peoples. Today, Arizona is home to 22 federally recognized tribes, with Tucson being home to the O'odham and the Yaqui. The university strives to build sustainable relationships with sovereign Native Nations and Indigenous communities through education offerings, partnerships, and community service.

This manuscript benefited from grammar checking and proofreading using ChatGPT \citep{openai_chatgpt_2023}.

\end{acknowledgments}

\begin{contribution}
YZ did the analysis and wrote this paper. EE, XF, and FS contributed to the observation design. GDB, YF, JMH, XJ and MP contributed to initial discussions and methodologies. CC and HC contributed to the theoretical interpretation. ACE and MY contributed to the primary observation as well as the EIGER observations. All authors contributed to discussions and interpretation of the results.
\end{contribution}

\vspace{5mm}
\facilities{JWST, MAST}

\software{
{\tt astropy} \citep{astropy_collaboration_astropy_2022},
{\tt JWST Calibration Pipeline} \citep{bushouse_jwst_2022}
}

\appendix
\renewcommand{\thefigure}{A\arabic{figure}}
\setcounter{figure}{0}

\section{Robustness Tests}
\label{sec:robustness}

\begin{figure*}[!ht]
    \centering
    \gridline{
        \fig{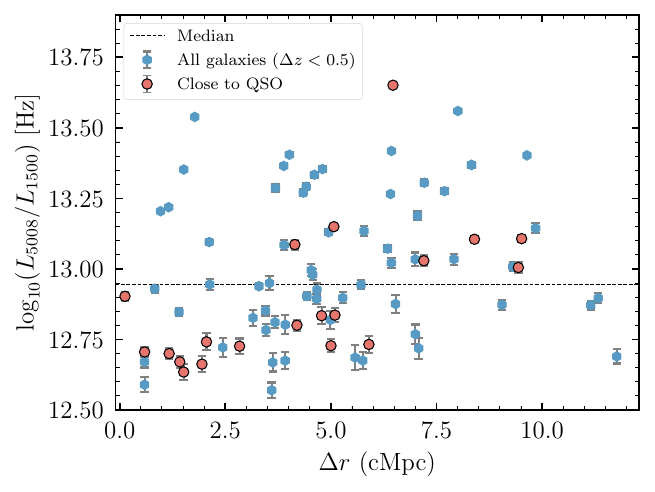}{0.49\linewidth}{(a)}
        \fig{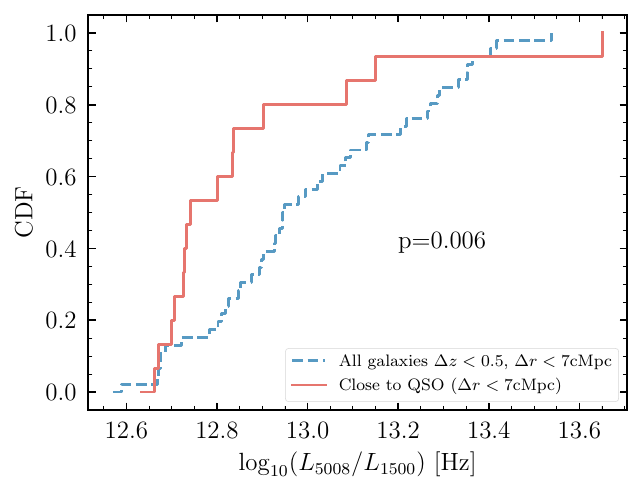}{0.49\textwidth}{(b)}
    }
    \gridline{
        \fig{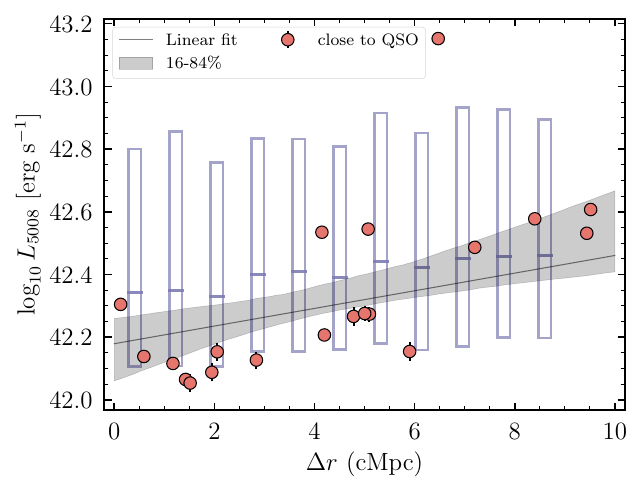}{0.49\linewidth}{(c)}
        \fig{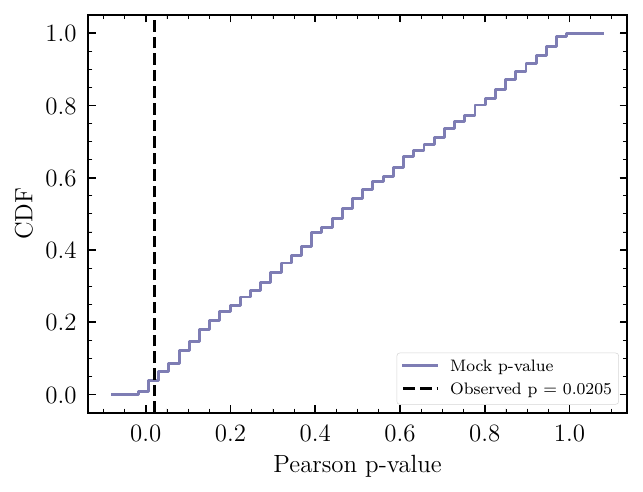}{0.49\linewidth}{(d)}
    }
    \caption{
Robustness tests of the suppression of $L_{5008}$ around J0100$+$2802.
(a) $\log_{10}(L_{5008}/L_{1500})$ as a function of projected transverse distance $\Delta r$ (cMpc). Red circles are galaxies near the quasar redshift; blue squares include all galaxies within $\Delta z<0.5$ of the quasar. The horizontal dashed line marks the median of the full sample.
(b) Cumulative distributions of $\log_{10}(L_{5008}/L_{1500})$ for galaxies within $\Delta r<7$\,cMpc (red: near $z_{\rm QSO}$; blue: all galaxies). A two-sample K--S test gives $p=0.006$.
(c) [O\,\textsc{iii}] luminosity versus $\Delta r$ for the near-$z_{\rm QSO}$ sample. The black line and gray band show the best-fit linear relation with 16--84\% bootstrap intervals. Purple boxes show the forward-modeled expectation from the \citet{matthee_eiger_2023} luminosity function convolved with survey completeness, plotted as per-bin medians with 16--84\% ranges (independent across bins).
(d) Distribution of $p$-values from linear fits to 1000 mock realizations (purple CDF). The vertical dashed line marks the observed correlation ($p=0.0205$), which lies in the $\sim2$\% tail of the null distribution, indicating that completeness variations alone is unlikely to reproduce the observed trend.
}
    \label{fig:app}
\end{figure*}

An important concern is whether the apparent suppression of [O\,\textsc{iii}] emission near J0100$+$2802 could be driven by variations in survey depth across the field. In this section we test the robustness of our result against such observational biases. Because the jump in $L_{5008}/L_{1500}$ in Figure~\ref{fig:trend} coincides with the transition from EIGER to SAPPHIRES coverage, we restrict this test to the EIGER data and to $\Delta r < 7$\,cMpc.

First, we note that the sensitivity is higher toward the field center (Figure~\ref{fig:trend}). If observational bias dominated, we would expect to recover both low- and high-$L_{5008}/L_{1500}$ galaxies close to the quasar. Instead, the opposite is observed. As shown in Figure~\ref{fig:app}(a), when selecting all galaxies within $|\Delta z|<0.5$ of the quasar, there exists a substantial population with high $L_{5008}/L_{1500}$ ratios. However, within $\Delta r < 7$\,cMpc of the quasar (red points), this high-ratio population is absent. The cumulative distributions in Figure~\ref{fig:app}(b) further highlight this effect: the close-to-QSO sample is significantly shifted toward lower $\log_{10}(L_{5008}/L_{1500})$ compared to the rest of the galaxies within the same redshift window (blue points). A two-sample K--S test yields $p=0.006$, confirming that the observed deficit of high $L_{5008}/L_{1500}$ galaxies near the quasar is statistically significant ($p<0.05$).

In the main text (Section~\ref{sec:results}), we downsampled the galaxy sample to enforce uniform completeness. Here we carry out a more sophisticated test by explicitly forward modeling the survey selection function. We convolve the \citet{matthee_eiger_2023} [O\,\textsc{iii}] luminosity function with the two-dimensional completeness maps as a function of source luminosity, redshift, and position, thereby accounting for both the higher central sensitivity and geometric coverage variations. The resulting mock catalogs predict the distribution of detected [O\,\textsc{iii}] emitters under the null hypothesis of no environmental or quasar-induced suppression. As shown in Figure~\ref{fig:app}(c), the forward-modeled median $L_{5008}$ values remain approximately flat with radius, with at most a weak increase ($<0.2$\,dex) at large $\Delta r$. The distribution of slopes and $p$-values from 1000 mock realizations indicates that the probability of reproducing the observed positive correlation is less than 2\% (Figure~\ref{fig:app}(d)).

Together, these tests demonstrate that the suppression of [O\,\textsc{iii}] luminosity near J0100$+$2802 is unlikely to be an artifact of varying sensitivity across the field. Instead, the deficit of bright [O\,\textsc{iii}] emitters close to the quasar is likely genuine and consistent with quasar-driven radiative feedback.

\bibliographystyle{aasjournalv7}



\end{document}